\documentclass[runningheads, envcountsame, a4paper]{llncs}
\usepackage{amssymb}
\usepackage{amsmath}		
\usepackage{amsthm}		%
\usepackage{enumerate} 		%
\usepackage{tikz}		%
\usetikzlibrary{automata}	
\usepackage{bbm}		%

\usepackage{MnSymbol} 		%
\usepackage{thmtools}		%
\usepackage{microtype}		%
\DeclareMathOperator{\acc}{acc}
\DeclareMathOperator{\wt}{wt}
\DeclareMathOperator{\Val}{Val}
\DeclareMathOperator{\free}{free}
\DeclareMathOperator{\const}{const}
\DeclareMathOperator{\call}{call}
\DeclareMathOperator{\ret}{ret}
\DeclareMathOperator{\pcall}{pcall}
\DeclareMathOperator{\pret}{pret}
\DeclareMathOperator{\Lab}{Lab}
\begin{document}
\spnewtheorem{Fuller}{Proposition}{\bfseries}{\itshape}	%
\spnewtheorem{Satz}{Proposition}{\bfseries}{\itshape}
\spnewtheorem{Folgerung}[Satz]{Corollary}{\bfseries}{\itshape}
\spnewtheorem{Theorem}[Satz]{Theorem}{\bfseries}{\itshape}
\spnewtheorem{Lemma}[Satz]{Lemma}{\bfseries}{\itshape}
\spnewtheorem{Def}[Satz]{Definition}{\bfseries}{\itshape}
\title{Weighted Automata and Logics for \\ Infinite Nested Words}
\author{Manfred Droste \and Stefan D\"uck\thanks{supported by Deutsche Forschungsgemeinschaft (DFG), project DR 202/11-1 and Graduiertenkolleg 1763 (QuantLA)}}
\institute{Institut f\"ur Informatik, University Leipzig, D-04109 Leipzig, Germany
\email{$\{$droste,dueck$\}$@informatik.uni-leipzig.de}
}
\authorrunning{M. Droste and S. D\"uck}
\maketitle
\begin{abstract}
Nested words introduced by Alur and Madhusudan are used to capture structures with both linear and hierarchical order, e.g. XML documents, without losing valuable closure properties. Furthermore, Alur and Madhusudan introduced automata and equivalent logics for both finite and infinite nested words, thus extending B\"uchi's theorem to nested words. Recently, average and discounted computations
of weights in quantitative systems found much interest.
Here, we will introduce and investigate weighted automata models
and weighted MSO logics for infinite nested words.
As weight structures we consider valuation monoids
which incorporate average and discounted computations of weights
as well as the classical semirings. We show that under suitable
assumptions, two resp. three fragments of our weighted logics
can be transformed into each other. Moreover, we show that
the logic fragments have the same expressive power as
weighted nested word automata.
\end{abstract}
	\section{Introduction}
Nested words, introduced by Alur and Madhusudan \cite{AM},
capture models with both a natural sequence of positions
and an hierarchical nesting of these positions.
Prominent examples include XML documents and executions of
recursively structured programs. Automata on nested words,
logical specifications, and corresponding languages of
nested words have been intensively studied, see \cite{AAB}, \cite{AM}, \cite{LMS}.
Recently, there has been much interest in quantitative features
for the specification and analysis of systems. Quantitative
automata modeling the long-time average or discounted behavior
of systems were investigated by Chatterjee, Doyen, and
Henzinger \cite{CDH}, \cite{CDH2}.
It is the goal of this paper to present quantitative logics for
such quantitative automata on nested words.

The connection between MSO logic and automata due to
B\"uchi, Elgot, and Trakhenbrot \cite{Bue}, \cite{Elg}, \cite{Tra} has proven most
fruitful. Weighted automata over semirings (like $(\mathbb{N},+,\cdot,0,1))$
were already investigated by Sch\"utzenberger \cite{Sch} and
soon developed a flourishing theory, cf. the books
\cite{BR}, \cite{Eil}, \cite{KS}, \cite{SS} and the recent handbook \cite{DKV}.
However, an expressively equivalent weighted MSO logic was
developed only recently \cite{DG}. This was extended to
semiring-weighted automata and logics over finite nested words
in \cite{Ma}, and further to strong bimonoids as weight structures
in \cite{DP}. For quantitative automata and logics, incorporating
average and discounting computations of weights over words,
such an equivalence was given in \cite{DM}.

In this paper, we will investigate quantitative nested word
automata and suitable quantitative MSO logics. We will
concentrate on infinite nested words, although our
results also hold for finite nested words. We employ the stair
Muller nested word automata of \cite{AM}, \cite{LMS}, since these
can be determinized without losing expressive power. As weight structures
we take the valuation monoids of \cite{DM}. These include
infinite products as in totally complete semirings \cite{DR}, but also
computations of long-time averages or discountings of weights. As example for such a setting
we give the calculation of the long-time ratio of
bracket-free positions in prefixes of an infinite nested word.
As our first main result, we show that under suitable assumptions
on the valuation monoid $D$, two resp. three versions of our
weighted MSO logic have the same expressive power.
In particular, if $D$ is commutative, then any weighted MSO-formula is equivalent
to one in which conjunctions occur only between 'classical'
boolean formulas and constants.
In contrast
to \cite{DM}, our proof uses direct conversions of the formulas
and thus has much better complexity than using the
automata-theoretic constructions of \cite{DM}.
These conversions are new even for the case of weighted logics on words.

In our second main result, we show under suitable assumptions
on the valuation monoid that our weighted MSO logics have
the same expressive power as weighted nested automata.
These assumptions on the valuation monoid are satisfied by
long-time average resp. discounted computations of weights;
therefore our results apply to these settings.
All our constructions of automata from formulas or conversely
are effective.
 	\section{Automata and Logics for Nested $\omega$-Words}
\label{chapnw}
In this section we describe basic background for classical (unweighted) automata and logics on nested-$\omega$-words. We denote by $\Sigma$ an alphabet
and by $\Sigma^\omega$ the set of all $\omega$-words over $\Sigma$. $\mathbb{N}$ is the set of all natural numbers without zero. For a binary relation $R$, we denote with $R(x,y)$ that $(x,y) \in R$.
\begin{Def}
	 A \emph{matching relation} $\nu$ over $\mathbb{N}$ is a subset of $(\{-\infty\} \cup \mathbb{N}) \times 	(\mathbb{N} \cup\{ \infty \})$ such that:
	\begin{enumerate}[\quad(i)]
		\item $\nu(i,j) \Rightarrow i<j$, 
		\item $\forall i \in \mathbb{N}: |\{j:\nu(i,j)\}| \le 1 \wedge \ |\{j:\nu(j,i)\}| \le 1$, 
		\item $\nu(i,j) \wedge \nu(i',j') \wedge i < i' \Rightarrow j < i' \vee j > j'$, 
		\item $(-\infty,\infty) \notin \nu$.
	\end{enumerate}
	A \emph{nested $\omega$-word} $\mathit{nw}$ over $\Sigma$ is a pair $(w,\nu)=(a_1a_2...,\nu)$ where $w=a_1a_2...$ is an $\omega$-word over $\Sigma$ and $\nu$ is a matching relation over $\mathbb{N}$.
	We denote by $\mathit{NW^\omega}(\Sigma)$ the set of all nested $\omega$-words over $\Sigma$ and we call every subset of $\mathit{NW^\omega}(\Sigma)$ a \emph{language of nested $\omega$-words}.
\end{Def}
	If $\nu(i,j)$ holds, we call $i$ a \emph {call position} and $j$ a \emph{return position}. In case of $j=\infty$, $i$ is a \emph{pending call} otherwise a \emph{matched call}. In case of $i=-\infty$, $j$ is a \emph{pending return} otherwise a \emph{matched return}. If $i$ is neither call nor return, then we say $i$ is an \emph{internal}. 
\begin{Def}
	A \emph{deterministic stair Muller nested word automaton (sMNWA)} over $\Sigma$ is a quadruple $\mathcal{A}=(Q,q_0,\delta,\mathfrak{F})$, where $\delta=(\delta_{\call},\delta_{\mathrm{int}},\delta_{\ret})$, consisting of:
\begin{itemize}
		\item a finite set of states $Q$,
		\item an initial state $q_0 \in Q$,
		\item a set $\mathfrak{F} \subseteq 2^Q$ of accepting sets of states,
		\item the transition functions $\delta_{\call},\delta_{\mathrm{int}}:Q \times \Sigma \rightarrow Q$,
 		\item the transition function $\delta_{\ret}:Q \times Q \times \Sigma \rightarrow Q$. 
	\end{itemize}	
\end{Def}
A \emph{run} $r$ of the sMNWA $\mathcal{A}$ on the nested $\omega$-word $\mathit{nw}=(a_1a_2...,\nu$) is an infinite sequence of states $r=(q_0,q_1,...)$ 
where $q_i \in Q$ for each $i \in \mathbb{N}$ 
and $q_0$ is the inital state of $\mathcal{A}$ such that for each $i \in \mathbb{N}$ the following holds:
\begin{align*}
 	\left \{ \begin{array}{ll}	
		\delta_{\call}(q_{i-1},a_i)=q_i 	&, \text{ if } \nu(i,j) \text{ for some } j>i \text{ (or $j=\infty$)} \\
		\delta_{\mathrm{int}}(q_{i-1},a_i)=q_i 	&, \text{ if $i$ is an internal } \\
		\delta_{\ret}(q_{i-1},q_{j-1},a_i)=q_i &, \text{ if } \nu(j,i) \text{ for some } 1 \le j<i \\
		\delta_{\ret}(q_{i-1},q_0,a_i)=q_i &, \text{ if } \nu(-\infty,i)\enspace.
	\end{array} \right .
\end{align*}
	 We call $i \in \mathbb{N}$ a \emph{top-level position} if there exist no positions $j,k \in \mathbb{N}$ with $j <i < k$ and $\nu(j,k)$. 
We define \begin{align*} Q^t_\infty(r)=\{q \in Q~|~q=q_i \mbox{ for infinitely many top-level positions }i\}\enspace. \end{align*} A run $r$ of an sMNWA is \emph{accepted} if %
$Q^t_\infty(r) \in \mathfrak{F}$.
An sMNWA $\mathcal{A}$ \emph{accepts} the nested $\omega$-word $\mathit{nw}$ if there is an accepted run of $\mathcal{A}$ on $\mathit{nw}$. We denote with $L(\mathcal{A})$ the set of all accepted nested $\omega$-words of $\mathcal{A}$. We call a language $L$ of nested $\omega$-words \emph{regular} if there is an sMNWA $\mathcal{A}$ with $L(\mathcal{A})=L$.
\par
Alur and Madhusudan \cite{AM} considered nondeterministic B\"uchi NWA and nondeterministic Muller NWA. They showed that the deterministic versions of these automata have strictly less expressive power than the nondeterministic automata. However, refering to L\"oding, Madhusudan and Serre \cite{LMS}, Alur and Madhusudan stated that deterministic stair Muller NWA have the same expressive power as their nondeterministic versions as well as nondeterministic B\"uchi NWA.
Moreover, the class of regular languages of nested-$\omega$-words is closed under union, intersection and complement (\cite{AM}).
\begin{Def}
	 The %
monadic second order logic for nested words %
$\mathit{MSO}(\mathit{NW}(\Sigma))$ contains exactly all %
formulas $\varphi$ which are given by the following syntax:
	\begin{align*}
	\varphi&::=\Lab_a(x)\, |\, \call(x)\, |\, \ret(x)\, |\, x \le y\, |\, \nu(x,y)\, |\, x \in X\, |\, \neg \varphi\, |\, \varphi \vee \varphi\, |\, \exists x. \varphi\, |\, \exists X. \varphi
	\end{align*}
where $a \in \Sigma$ and $x,y$ are first order variables and $X$ 
is a second order variable.
\end{Def}
The semantics of these formulas is given in a natural way%
, cf. \cite{AM}. %
Later we give a full definition of the semantics of \emph{weighted} MSO-formulas. 
We call $\varphi$ a \emph{sentence} if $\varphi$ contains no free variables. If $\varphi$ is a sentence, then
$L(\varphi)=\{\mathit{nw} \in \mathit{NW^\omega}(\Sigma)~|~\mathit{nw} \models \varphi\}$
 is \emph{the language defined by} $\varphi$.
\begin{Theorem}[Alur, Madhusudan \cite{AM}] \label{regMSOD} %
	Let $L$ be a language of nested $\omega$-words over $\Sigma$. Then $L$ is regular if and only if $L$ is definable by some $\mathit{MSO}(\mathit{NW}(\Sigma))$-sentence $\varphi$. 
\end{Theorem}
	\section{Weighted Stair Muller Nested Word Automata} \label{kapomega} %
In this section, we introduce weighted versions of stair Muller nested word automata. As weight structures, we will employ $\omega$-valuation monoids introduced in \cite{DM}. We recall the definitions.
\par
A monoid $(D,+,0)$ is \emph{complete} if it has infinitary sum operations
	$\sum_I:D^I \rightarrow D$
	for any index set $I$ such that
	\begin{itemize}
		\item $\sum_{i \in \emptyset} d_i = 0$,
		 $\sum_{i \in \{k\}} d_i = d_k$,
		 $\sum_{i \in \{j,k\}} d_i = d_j+d_k$ for $j \neq k$,
		\item $\sum_{j \in J}(\sum_{i \in I_j} d_i ) = \sum_{i \in I} d_i \text{ if } \bigcup_{j \in J} I_j = I \text{ and } I_j \cap I_k = \emptyset \text{ for } j \neq k$.
	\end{itemize}
	Note that in every complete monoid the operation $+$ is commutative.
We let $D^\omega$ comprise all infinite sequences of elements of $D$.
\begin{Def}[Droste, Meinecke \cite{DM}]
	An \emph{$\omega$-valuation monoid} $(D,+,\Val^\omega,0)$ is a complete monoid $(D,+,0)$ equipped with an \emph{$\omega$-valuation function}
$\Val^\omega:D^\omega \rightarrow D$ with $\Val^\omega((d_i)_{i \in \mathbb{N}})=0$ if $d_i=0$ for some $i \in \mathbb{N}$.%
\par
	A \emph{product $\omega$-valuation monoid} $(D,+,\Val^\omega, \diamond,0,1)$ (short \emph{$\omega$-pv-monoid}) is an $\omega$-valuation monoid $(D,+,\Val^\omega,0)$ with a constant $1 \in D$ and an operation $\diamond:D^2 \rightarrow D$ satisfying $\Val^\omega(1^\omega)=1,~0 \diamond d= d \diamond 0 = 0 \text{ and } 1 \diamond d = d \diamond 1 = d \text{ for all } d \in D$.
\end{Def}
	Let $(D,+,\Val^\omega, \diamond,0,1)$ be an $\omega$-pv-monoid. $D$ is called \emph{associative} resp. \emph{commutative} if $\diamond$ is associative resp. commutative. 
	$D$ is \emph{left-$+$-distributive} if for all $d \in D$, for any index set $I$ and $(d_i)_{i \in I}\in D^I$:
		\begin{align*}d\diamond \sum_{i \in I} d_i=\sum_{i \in I} (d \diamond d_i)\enspace. \end{align*}
	\emph{Right-${+}$-distributivity} is defined analogously. We call $D$ \emph{$+$-distributive} if $D$ is left- and right-$+$-distributive. 
	$D$ is \emph{left-$\text{Val}^\omega$-distributive} if for all $d \in D$ and $(d_i)_{i \in \mathbb{N}}\in D^\omega$:
		\begin{align*}d \diamond \Val^\omega((d_i)_{i \in \mathbb{N}})=\Val^\omega((d \diamond d_i)_{i \in \mathbb{N}} ) \enspace. \end{align*}
	$D$ is \emph{left-multiplicative} if for all $d \in D$ and $(d_i)_{i \in \mathbb{N}}\in D^\omega$:
		\begin{align*}d \diamond \Val^\omega((d_i)_{i \in \mathbb{N}})=\Val^\omega(d \diamond d_1,(d_i)_{i \ge 2}) \enspace. \end{align*}
	$D$ is called \emph{conditionally commutative}, if for all $(d_i)_{i \in \mathbb{N}}$, $(d_i')_{i \in \mathbb{N}}\in D^\omega$ with $d_i \diamond d_j' = d_j' \diamond d_i$ for all $j < i$, the following holds:
\begin{align*} \Val^\omega((d_i)_{i \in \mathbb{N}}) \diamond \Val^\omega((d_i')_{i \in \mathbb{N}})=\Val^\omega((d_i\diamond d_i')_{i \in \mathbb{N}})\enspace. \end{align*}
	We call $D$ \emph{left-distributive} if $D$ is left-$+$-distributive and, additionally, left-$\Val^\omega$-distributive or left-multiplicative. 
	If $D$ is $+$-distributive and associative, then $(D,+, \diamond,0,1)$ is a complete semiring and we call $(D,+,\Val^\omega, \diamond,0,1)$ an \emph{$\omega$-valuation semiring}. 
	A \emph{cc-$\omega$-valuation semiring} is an $\omega$-valuation semiring $D$ which is conditionally commutative and left-distributive.
\begin{example}[\cite{DM}] We set $\bar{\mathbb{R}}=\mathbb{R} \cup \{-\infty, \infty\}$ and $-\infty +\infty = -\infty$. We let %
\begin{align*}
	 (D_1,+,\Val^\omega, \diamond,0,1)&=(\bar{\mathbb{R}},\sup,\text{lim avg},+,-\infty,0), \\
	\text{ where } \hspace{2cm}	 \text{lim avg}((d_i)_{i \in \mathbb{N}})&=\liminf_{n \rightarrow \infty}\frac{1}{n}\sum_{i=1}^{n}d_i\enspace. \hspace{3cm}
\end{align*}
Let $0 < \lambda < 1$ and $\bar{\mathbb{R}}_{+}=\{x\in \bar{\mathbb{R}}~|~x \ge 0\} \cup \{-\infty\}$. We put %
\begin{align*}
	 (D_2,+,\Val^\omega, \diamond,0,1)&=(\bar{\mathbb{R}}_{+},\sup,\mathrm{disc}_\lambda,+,-\infty,0), \\
	\text{where } \hspace{2cm} \mathrm{disc}_\lambda((d_i)_{i \in \mathbb{N}})&=\lim_{n \rightarrow \infty}\sum_{i=1}^{n}\lambda^{i-1}d_i\enspace. \hspace{3cm}
\end{align*}
Then $D_1$ is a left-$+$-distributive and %
left-$\Val^\omega$-distributive $\omega$-valuation monoid but not conditionally commutative. Furthermore, $D_2$ is a left-multiplicative cc-$\omega$-valuation semiring. 
\end{example}
\begin{Def} \label{defwsMNWA}
	A \emph{weighted stair Muller nested word automaton (wsMNWA)} $\mathcal{A}=(Q,I,\delta,\mathfrak{F})$, where $\delta=(\delta_{\call},\delta_{\mathrm{int}},\delta_{\ret})$, over the alphabet $\Sigma$ and the $\omega$-valuation monoid $(D,+,\Val^\omega,0)$ consists of:
	\begin{itemize}
		\item a finite set of states $Q$,
		\item a set $I \subseteq Q$ of initial states,
		\item a set $\mathfrak{F} \subseteq 2^Q$ of accepting sets of states,
		\item the weight functions $\delta_{\call},\delta_{\mathrm{int}}:Q \times \Sigma \times Q \rightarrow D$,
 		\item the weight function $\delta_{\ret}:Q \times Q \times \Sigma \times Q \rightarrow D$.
	\end{itemize}
\end{Def}
A \emph{run} $r$ of the wsMNWA $\mathcal{A}$ on the nested $\omega$-word $\mathit{nw}=(a_1a_2...,\nu$) is an infinite sequence of states $r=(q_0,q_1,...)$. %
We denote with $\mathit{wt}_\mathcal{A}(r,\mathit{nw},i)$ %
the weight of the transition of $r$ used at position $i \in \mathbb{N}$, defined as follows
\begin{align}
	 \wt_\mathcal{A}(r,\mathit{nw},i)&=\left \{ \begin{array}{ll}\delta_{\call}(q_{i-1},a_i,q_i) &, \text{ if } \nu(i,j) \text{ for some } j>i\\
					\delta_{\mathrm{int}}(q_{i-1,}a_i,q_i) &, \text{ if $i$ is an internal} \\
	\label{autgewichte}		\delta_{\ret}(q_{i-1},q_{j-1},a_i,q_i) &, \text{ if } \nu(j,i) \text{ for some } 1 \le j<i \\
					\delta_{\ret}(q_{i-1},q_{I},a_i,q_i) &, \text{ if } \nu(-\infty,i) \text{ for some } q_I \in I\enspace.
					\end{array} \right .
\end{align}
Then we define the \emph{weight} $\mathit{wt}_\mathcal{A}(r,\mathit{nw})$ %
\emph{of $r$ on $\mathit{nw}$} by letting
\begin{align*}
	\wt_\mathcal{A}(r,\mathit{nw})&=\Val^\omega((\wt_\mathcal{A}(r,\mathit{nw},i))_{i \in \mathbb{N} })\enspace. %
\end{align*}
We define top-level positions and the set $Q^t_\infty(r)$ as before. A run $r$ is \emph{accepted} if %
$q_0 \in I$ and $Q^t_\infty(r) \in \mathfrak{F}$. %
We denote with $\mathit{acc}(\mathcal{A})$ the set of all accepted runs in $\mathcal{A}$. We define the \emph{behavior of the automaton $\mathcal{A}$} as the function $\lVert \mathcal{A} \rVert: \mathit{NW^\omega}(\Sigma) \rightarrow D$ given by (where as usual, empty sums are defined to be $0$) 
\begin{align*}
\notag	\lVert \mathcal{A} \rVert(\mathit{nw})&=\sum_{r \in \acc(\mathcal{A})} \wt_\mathcal{A}(r,\mathit{nw})\\
	 	 &=\sum_{r \in \acc(\mathcal{A})}\Val^\omega((\wt_\mathcal{A}(r,\mathit{nw},i))_{i \in \mathbb{N} })\enspace.
\end{align*}
\par
	We call every function $S:\mathit{NW^\omega}(\Sigma) \rightarrow D$ a \emph{nested $\omega$-word series} (short: \emph{series}).
	We call a series $S$ \emph{regular} if there exists an automaton $\mathcal{A}$ with $\lVert \mathcal{A} \rVert=S$.
\begin{example} \label{example:series}
	Within the following example we call a position $i$ of a nested $\omega$-word $\mathit{nw}=(w,\nu)$ \emph{bracketfree} if there are no positions $j,k \in (\mathbb{N}\cup \{-\infty,\infty\})$ with $j< i < k$ and $\nu(j,k)$. %
	This requirement is stronger than $i$ being a top-level position because it contains $-\infty$ and $\infty$ thus also banning $i$ being in the scope of pending calls and pending returns. Only for well-matched nested $\omega$-words, i.e. nested $\omega$-words without pending edges, the two properties coincide. \par
	We consider the series $S$ assigning to every nested $\omega$-word $\mathit{nw}$ the greatest accumulation point of the ratio of bracketfree positions in finite prefixes of $\mathit{nw}$. %
	\par
	To model $S$ we use the $\omega$-valuation monoid $D=(\bar{\mathbb{R}},\sup,\text{lim avg},-\infty)$. %
	If we want to analyze this property for well-matched nested $\omega$-words only, then automaton $\mathcal{A}_1$ given below recognizes $S$.
	In the general case including pending edges, automaton $\mathcal{A}_2$ recognizes $S$.
	Note that we denote the call transitions with $\langle \Sigma$ and the return transitions with $\Sigma \rangle /q $ where $q$ has to be the state where the last open call was encountered. The weights $1$ resp. $0$ are given in brackets. \par
~\\
\textbf{Automaton 1:} wsMNWA $\mathcal{A}_1$ with $\mathfrak{F_1}=\{\{q_0\}\}$
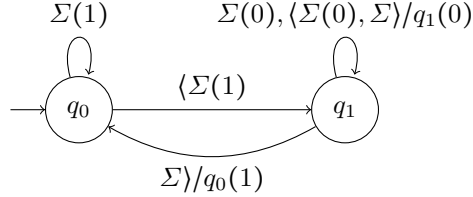
	\begin{center}
		\begin{tikzpicture}[->, auto, node distance = 3.5cm]
			\node [state, initial left, initial text = ] (A) {$q_0$};
			\node [state , right of = A] (B) {$q_1$};
			\path 	(A) edge [loop above] node {$ \Sigma (1)$} (A)
				(B) edge [loop above] node {$\Sigma(0), \langle \Sigma(0),\Sigma \rangle /q_1(0)$} (B)
				(A) edge node [above] {$\langle \Sigma (1)$} (B)
				(B) edge [bend left] node [below] {$\Sigma \rangle /q_0(1)$} (A);
		\end{tikzpicture}
	\end{center}
\textbf{Automaton 2:} wsMNWA $\mathcal{A}_2$ with $\mathfrak{F}_1 = \{\{q_2\},\{q_p\},\{q_2,q_p\},\{q_0,q_1\},\{q_0\},\{q_1\}\}$
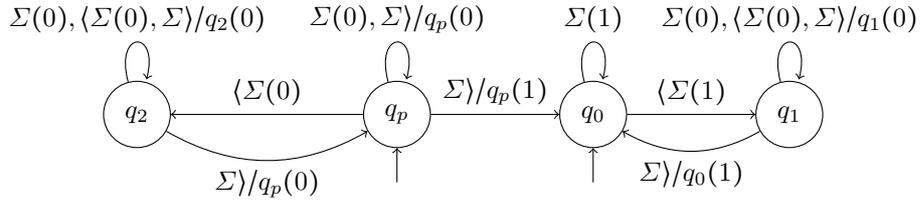
	\begin{center}
		\begin{tikzpicture}[->, auto, node distance = 2.58cm]
			\node [state, initial below, initial text = ] (P) {$q_p$};
			\node [state, initial below, initial text =, right of = P ] (A) {$q_0$};
			\node [state, left of = P, xshift=-0.84cm] (Q) {$q_2$};
			\node [state , right of = A] (B) {$q_1$};
			\path 	(A) edge [loop above] node {$ \Sigma (1)$} (A)
				(B) edge [loop above] node {$\Sigma(0), \langle \Sigma(0),\Sigma \rangle /q_1(0)$} (B)
				(P) edge [loop above] node {$\Sigma(0), \Sigma \rangle /q_p(0)$} (P)
				(P) edge node [above] {$ \Sigma \rangle /q_p(1)$} (A)
				(A) edge node [above] {$\langle \Sigma (1)$} (B)
				(B) edge [bend left] node [below] {$\Sigma \rangle /q_0(1)$} (A)
				(Q) edge [loop above] node {$\Sigma(0), \langle \Sigma(0),\Sigma \rangle /q_2(0)$} (Q)
				(P) edge node [above] {$\langle \Sigma (0)$} (Q)
				(Q) edge [bend right] node [below] {$\Sigma \rangle /q_p(0)$} (P);
		\end{tikzpicture}
	\end{center}
\end{example}
As usual, we extend the operation $+$ and $\diamond$ to series $S,T:\mathit{NW^\omega}(\Sigma) \rightarrow D$ by means of pointwise definitions as follows:
\begin{align*}
(S \star T)(\mathit{nw}) &= S(\mathit{nw}) \star T(\mathit{nw}) \mbox{ for each } \mathit{nw} \in \mathit{NW^\omega}(\Sigma), \star \in \{{+},{\diamond}\} \enspace .
\end{align*}
We let $d \in D$ also denote the constant series with value $d$, i.e. $\lVert d \rVert(\mathit{nw}) = d$ for each $\mathit{nw} \in \mathit{NW^\omega}(\Sigma)$.
For $L\subseteq \mathit{NW^\omega}(\Sigma)$, we define the \emph{characteristic series} $\mathbbm{1}_{L}:\mathit{NW^\omega}(\Sigma) \rightarrow D$ by letting $\mathbbm{1}_{L}(\mathit{nw})=1$ if $\mathit{nw} \in L$, and $\mathbbm{1}_{L}(\mathit{nw})=0$ otherwise.
We call a series $S$ a \emph{regular step function} if
\begin{align} \label{defrsf} S=\sum_{i=1}^k d_i \diamond \mathbbm{1}_{L_i}\enspace ,
\end{align}
where $L_i$ are regular languages of nested-$\omega$-words
	forming a partition of $\mathit{NW^\omega}(\Sigma)$ and $d_i \in D$ for each $i \in \{1,...,k\}$; so
	$S(\mathit{nw})=d_i$ iff $\mathit{nw} \in L_i$
	for each $i \in \{1,...,k\}$.
\par
An $\omega$-pv-monoid $D$ is \emph{regular} if for any alphabet $\Sigma$ we have: For each $d \in D$ there exists a wsMNWA $\mathcal{A}_d$ with $\lVert \mathcal{A}_d \rVert= d$. %
Analogously to Droste and Meinecke \cite{DM} we can show that every left-distributive $\omega$-pv-monoid is regular. %
\begin{Satz}
\label{rsfreg}
	Let $D$ be a regular $\omega$-pv-monoid. Then %
	each regular step function $S:\mathit{NW^\omega}(\Sigma) \rightarrow D$ is regular. Furthermore, the set of all regular step functions is closed under $+$ and $\diamond$.
\end{Satz}
Next we show that regular series are closed under projections. Consider a mapping $h:\Sigma \rightarrow \Gamma$ between two alphabets. Then $h$ extends uniquely to an homomorphism between $\Sigma^\omega$ and $\Gamma^\omega$, also denoted by $h$. Hence $h$ is length-preserving and we can extend $h$ to a function $h:\mathit{NW^\omega}(\Sigma) \rightarrow \mathit{NW^\omega}(\Gamma)$ by defining $h(\mathit{nw})=h(w,\nu)=(h(w),\nu)$ for each $\mathit{nw}\in \mathit{NW^\omega}(\Sigma)$.
Let $S:\mathit{NW^\omega}(\Sigma) \rightarrow D$ be a series. Then we define $h(S):\mathit{NW^\omega}(\Gamma) \rightarrow D$ for each $\mathit{nv}\in \mathit{NW^\omega}(\Gamma)$ by
\begin{align*}
	h(S)(\mathit{nv})&=\sum(S(\mathit{nw})~|~\mathit{nw} \in \mathit{NW^\omega}(\Sigma), h(\mathit{nw})=\mathit{nv}) \enspace.	
\end{align*}
\begin{Satz}
	\label{hom}
	Let $D$ be an $\omega$-valuation monoid, $S:\mathit{NW^\omega}(\Sigma) \rightarrow D$ regular and $h:\Sigma \rightarrow \Gamma$. Then $h(S):\mathit{NW^\omega}(\Gamma) \rightarrow D$ is regular.
\end{Satz}
	\section{Weighted MSO-Logic for Nested $\omega$-Words} \label{kapMSOD}
In this section, we will present different fragments of our weighted MSO logic, and we give our first main result on the equivalence of these fragments. 
In the following $D$ is always an $\omega$-pv-monoid. We combine ideas of Alur and Madhusudan \cite{AM}, Droste and Gastin \cite{DG}, and Bollig and Gastin \cite{BG}, 
and divide the syntax of the weighted logic into a boolean part and a weighted part.
\begin{Def}[Syntax]
	The weighted monadic second order logic for nested words $\mathit{MSO}(D, \mathit{NW}(\Sigma))$ is given by the following syntax
	\begin{align*}
	\beta&::=\Lab_a(x)\, |\, \call(x)\, |\, \ret(x)\, |\, x \le y\, |\, \nu(x,y)\, |\, x \in X\, |\, \neg \beta\, |\, \beta \wedge \beta\, |\, \forall x. \beta\, |\, \forall X. \beta \\
	\varphi&::=d~|~\beta~|~\varphi \vee \varphi~|~\varphi \wedge \varphi~|~\forall x. \varphi~|~\exists x. \varphi~|~\exists X. \varphi
	\end{align*}
where $d \in D$,~$a \in \Sigma$ and $x$, $y$, $X$ %
are first resp. second order variables. We call all formulas $\beta$ \emph{boolean formulas}.
\end{Def}
The set of all positions of $\mathit{nw} \in \mathit{NW^\omega}(\Sigma)$ is $\mathbb{N}$. %
Let $\varphi \in \mathit{MSO}(D, \mathit{NW}(\Sigma))$. We denote the set of free variables of $\varphi$ by $\free(\varphi)$. Let $\mathcal{V}$ be a finite set of variables containing $\free(\varphi)$. As usual, we define a \emph{$(\mathcal{V}, \mathit{nw})$-assignment} $\gamma$ as function assigning to every first order variable of $\mathcal{V}$ a position of $\mathit{nw}$ and to every second order variable a subset of positions of $\mathit{nw}$.
We let
$\gamma[x \rightarrow i]$ (resp. $\gamma[X \rightarrow I]$) be the $(\mathcal{V}\cup \{x\}, \mathit{nw})$-assignment (resp. $(\mathcal{V}\cup \{X\}, \mathit{nw})$)-assignment) mapping $x$ to $i$ (resp. $X$ to $I$) and equaling $\gamma$ anywhere else. \par
We encode a pair $(\mathit{nw},\gamma)$ %
as nested $\omega$-word %
as usual %
over the extended alphabet $\Sigma_\mathcal{V}=\Sigma \times \{0,1\}^\mathcal{V}$ with the same matching relation $\nu$ (cf. \cite{DG}, \cite{DP}). %
We call $(\mathit{nw},\sigma) \in \mathit{NW^\omega}(\Sigma_\mathcal{V})$ \emph{valid} if $\sigma$ emerges from a $(\mathcal{V}, \mathit{nw})$-assignment. Clearly the language $N_\mathcal{V}$ of all valid words is regular. %
\begin{Def}[Semantics]
\label{tab}
The %
semantics of $\varphi$ is a series %
$\lsem \varphi \rsem_\mathcal{V} : \mathit{NW^\omega}(\Sigma_\mathcal{V}) \rightarrow D$. If $(\mathit{nw},\sigma)$ is not valid, we set $\lsem \varphi \rsem_\mathcal{V}(\mathit{nw},\sigma)=0$. Otherwise we define \\ $\lsem \varphi \rsem_\mathcal{V}(\mathit{nw},\sigma)$ for $(\mathit{nw},\sigma)=((a_1a_2...,\nu),\sigma)$ inductively as follows:
	\begin{small}
	\begin{alignat*}{2}	
		\lsem \Lab_a(x) \rsem_\mathcal{V} (\mathit{nw},\sigma)&=
			\left \{\begin{array}{ll}	1 &,\text{if } a_{\sigma(x)}=a\\
							0 &, \text{otherwise,}	
				\end{array} \right . &
		\lsem \call(x) \rsem_\mathcal{V} (\mathit{nw},\sigma)&=
			\left \{\begin{array}{ll}	1 &,\text{if } \sigma(x) \text { is a call}\\
							0 &, \text{otherwise,}	
				\end{array} \right .\\
		\lsem \ret(x) \rsem_\mathcal{V} (\mathit{nw},\sigma)&=
			\left \{\begin{array}{ll}	1 &,\text{if } \sigma(x) \text { is a return} \\
							0 &, \text{otherwise,}	
				\end{array} \right . &
		\lsem x \le y \rsem_\mathcal{V} (\mathit{nw},\sigma)&=
			\left \{\begin{array}{ll}	1 &,\text{if } \sigma(x) \le \sigma(y) \\
							0 &, \text{otherwise,}	
				\end{array} \right . \\
		\lsem \nu(x,y) \rsem_\mathcal{V} (\mathit{nw},\sigma)&=
			\left \{\begin{array}{ll}	1 &,\text{if } \nu(\sigma(x),\sigma(y))\\
							0 &, \text{otherwise,}	
				\end{array} \right . &
		\lsem x \in X\rsem_\mathcal{V} (\mathit{nw},\sigma)&=
			\left \{\begin{array}{ll}	1 &,\text{if } \sigma(x) \in \sigma(X)\\
							0 &, \text{otherwise,}	
				\end{array} \right . \\
		\lsem \neg \beta \rsem_\mathcal{V} (\mathit{nw},\sigma)&=
			\left \{\begin{array}{ll}	1 &,\text{if } \lsem \beta \rsem_\mathcal{V} (\mathit{nw},\sigma)=0 \\
							0 &, \text{otherwise,}	
				\end{array} \right . &
		\lsem d \rsem_\mathcal{V} (\mathit{nw},\sigma)&= d \text{\quad for all } d\in D,
	\end{alignat*}
	\vspace{-16pt}				%
	\begin{align*}
		\lsem \varphi \vee \psi \rsem_\mathcal{V} (\mathit{nw},\sigma)&= \lsem \varphi \rsem_\mathcal{V} (\mathit{nw},\sigma) + \lsem \psi \rsem_\mathcal{V} (\mathit{nw},\sigma), \\
		\lsem \varphi \wedge \psi \rsem_\mathcal{V} (\mathit{nw},\sigma)&= \lsem \varphi \rsem_\mathcal{V} (\mathit{nw},\sigma) \diamond \lsem \psi \rsem_\mathcal{V} (\mathit{nw},\sigma),\\
		\lsem \exists x. \varphi \rsem_\mathcal{V} (\mathit{nw},\sigma)&= \sum_{i \in \mathbb{N}}( \lsem \varphi \rsem_{\mathcal{V}\cup \{x\}} (\mathit{nw},\sigma[x \rightarrow i])), \\
		\lsem \exists X. \varphi \rsem_\mathcal{V} (\mathit{nw},\sigma)&= \sum_{I \subseteq \mathbb{N}}( \lsem \varphi \rsem_{\mathcal{V}\cup \{X\}} (\mathit{nw},\sigma[X \rightarrow I])), \\
		\lsem \forall x. \varphi \rsem_\mathcal{V} (\mathit{nw},\sigma) &= \Val^\omega((\lsem \varphi \rsem_{\mathcal{V}\cup \{x\}} (\mathit{nw},\sigma [x \rightarrow i]))_{i \in \mathbb{N}}),\\
		\lsem \forall X. \beta \rsem_\mathcal{V} (\mathit{nw},\sigma)
		&=\left\{\begin{array}{ll}	1 &,\text{ if } \lsem \beta \rsem_{\mathcal{V}\cup \{X\}}(\mathit{nw},\sigma [X \rightarrow I]) =1~ \text{for all } I \subseteq \mathbb{N} \\
						0 &,\text{ otherwise\enspace. } \end{array}\right .
	\end{align*}
	\end{small}	%
\end{Def}
We write $\lsem \varphi \rsem$ for $\lsem \varphi \rsem_{\free(\varphi)}$, so
	$\lsem \varphi \rsem : \mathit{NW^\omega}(\Sigma_{\free(\varphi)}) \rightarrow D$.
If $\varphi$ contains no free variables, $\varphi$ is a \emph{sentence} and
	$\lsem \varphi \rsem : \mathit{NW^\omega}(\Sigma)\rightarrow D$.
\begin{example}
Continuing Example \ref{example:series} with $D=(\bar{\mathbb{R}},\sup,\text{lim avg},+,-\infty,0)$ %
we define %
\begin{align*}
&\mathrm{pcall}(x)=\call(x) \wedge \forall w.\neg\nu(x,w),\quad \pret(z)=\ret(z) \wedge \forall u.\neg\nu(u,z), \\
&\mathrm{bfr}(y)=\forall x \forall z.(\neg(x<y<z \wedge \nu(x,z)) \wedge \neg(x<y \wedge \pcall(x)) \wedge \neg(y<z \wedge \pret(z))) ,
\end{align*}
where $x<y<z=\neg(y\le x)\wedge\neg(z\le y)$. %
Then $\lsem \forall y.((\mathrm{bfr}(y)\wedge 1)\vee 0) \rsem = S =\lVert \mathcal{A}_2 \rVert$.
\end{example}
\par Analogously to \cite{DG} and \cite{DP} we can show:
\begin{Satz}
\label{prop:consistency}
	Let $\varphi \in \mathit{MSO}(D, \mathit{NW}(\Sigma))$ and let $\mathcal{V}$ be a finite set of variables with $\free(\varphi) \subseteq \mathcal{V}$. Then $\lsem \varphi \rsem_{\mathcal{V}}(\mathit{nw},\sigma)= \lsem \varphi \rsem(\mathit{nw},\sigma\restriction \free(\varphi))$ for each valid $(\mathit{nw},\sigma)\in \mathit{NW^\omega}(\Sigma_\mathcal{V})$. Furthermore, $\lsem \varphi \rsem$ is regular iff $\lsem \varphi \rsem_{\mathcal{V}}$ is regular. 
\end{Satz}
\par
Clearly, every boolean formula $\beta \in \mathit{MSO}(D, \mathit{NW}(\Sigma))$ can be interpreted as an unweighted MSO-formula $\psi \in \mathit{MSO}(\mathit{NW}(\Sigma))$ with $\lsem\beta \rsem=\mathbbm{1}_{L(\psi)}$, since $\lsem \beta \rsem$ only yields the values $0$ and $1$.
Conversely, for every formula $\psi \in \mathit{MSO}(\mathit{NW}(\Sigma))$ there exists a boolean MSO-formula $\beta \in \mathit{MSO}(D, \mathit{NW}(\Sigma))$ with $\lsem \beta \rsem=\mathbbm{1}_{L(\psi)}$, since we can replace disjunctions by conjunctions and negations and we can replace existential quantifiers by universal quantifiers and negations.
\par
In order to obtain a B\"uchi-like theorem (as Theorem \ref{main} below) for weighted automata on finite words, it is necessary to restrict the weighted MSO logic (cf. \cite{DG}). Therefore we introduce and study suitable fragments of $\mathit{MSO}(D, \mathit{NW}(\Sigma))$ as in the following.
\begin{Def}
	The \emph{set of almost boolean formulas} is the smallest set of all formulas of $\mathit{MSO}(D, \mathit{NW}(\Sigma))$ containing all constants $d \in D$ and all boolean formulas, which is closed under disjunction and conjunction.
\end{Def}
\begin{Satz}
	\label{almboolrsf}%
	 \begin{enumerate}[\quad (a)]
	\item If $\varphi \in \mathit{MSO}(D, \mathit{NW}(\Sigma))$ is an almost boolean formula, then $\lsem \varphi \rsem$ is a regular step function.
	\item For every regular step function $S: \mathit{NW^\omega}(\Sigma)\rightarrow D$, there exists an almost boolean sentence $\varphi$ with $S=\lsem \varphi \rsem$.
	\end{enumerate}
\end{Satz}
\begin{Def}%
	Let $\varphi \in \mathit{MSO}(D, \mathit{NW}(\Sigma))$. We denote by $\const(\varphi)$ the set of all elements of $D$ occurring in $\varphi$. %
We call $\varphi$
	\begin{enumerate}
		\item \emph{strongly-$\wedge$-restricted} if for all subformulas $\psi \wedge \theta$ of $\varphi$: \\
		Either $\psi$ and $\theta$ are almost boolean or $\psi$ is boolean or $\theta$ is boolean.
		\item \emph{$\wedge$-restricted} if for all subformulas $\psi \wedge \theta$ of $\varphi$: \\
 		Either $\psi$ is almost boolean or $\theta$ is boolean.
		\item \emph{commutatively-$\wedge$-restricted} if for all subformulas $\psi \wedge \theta$ of $\varphi$: \\
 		Either $\const(\psi)$ and $\const(\theta)$ commute or $\psi$ is almost boolean. %
		\item \emph{$\forall$-restricted} if for all subformulas $\forall x. \psi$ of $\varphi$: %
		 $\psi$ is almost boolean.
	\end{enumerate}
\end{Def}
We call a formula of $\mathit{MSO}(D, \mathit{NW}(\Sigma))$ \emph{syntactically restricted} if it is both $\forall$-restricted and strongly-$\wedge$-restricted. %
Note that every subformula of a syntactically restricted formula is syntactically restricted itself.
~\par
Now we show that under suitable assumptions on the $\omega$-pv-monoid $D$, particular classes of $\mathit{MSO}(D, \mathit{NW}(\Sigma))$-formulas have the same expressive power. In \cite{DM} these equivalences (for unnested words) followed from the main result and thus needed constructions of automata. Here we show the equivalence of the logic fragments directly.
\begin{Theorem}
	\label{thm:restrict}
	\begin{enumerate}[\quad(a)]
		\item Let $D$ be left-distributive and $\varphi \in \mathit{MSO}(D, \mathit{NW}(\Sigma))$ be $\wedge$-restricted. Then there exists a strongly-$\wedge$-restricted formula \\ $\varphi' \in \mathit{MSO}(D, \mathit{NW}(\Sigma))$ with $\lsem \varphi \rsem=\lsem \varphi' \rsem$. Moreover, if $\varphi$ is also $\forall$-restricted, then $\varphi'$ can also be chosen to be $\forall$-restricted.
		\item Let $D$ be a cc-$\omega$-valuation semiring and let $\varphi \in \mathit{MSO}(D, \mathit{NW}(\Sigma))$ be commutatively-$\wedge$-restricted. Then there exists a strongly-$\wedge$-restricted formula
$\varphi' \in \mathit{MSO}(D, \mathit{NW}(\Sigma))$ with $\lsem \varphi \rsem=\lsem \varphi' \rsem$. Moreover, if $\varphi$ is also $\forall$-restricted, then $\varphi'$ can also be chosen to be $\forall$-restricted.
	\end{enumerate}
\end{Theorem}
\begin{proof}[Proof (sketch)]
	We use an induction on the structure of $\varphi$. The interesting case is $\varphi=\psi \wedge \theta$ and $\psi$ is almost boolean.
By induction we can assume that $\psi$ and $\theta$ are strongly-$\wedge$-restricted (and resp. $\forall$-restricted).
As an example,
we consider the case of the universal quantification in $(a)$ as follows.
Assume $\theta=\forall x.\theta_1$ and $\psi$ does not contain $x$.
By the induction hypothesis, we obtain a strongly-$\wedge$-restricted formula $\varphi_1$ such that $\lsem \varphi_1 \rsem = \lsem \psi \wedge \theta_1 \rsem$.
\par 
First let $D$ be left-$\Val^\omega$-distributive. Using this assumption at equation *, we get for $\mathcal{V}=\free(\psi) \cup \free(\forall x.\theta_1)$ and each $(\mathit{nw},\sigma)\in \mathit{NW^\omega}(\Sigma_\mathcal{V})$:
\begin{align*} \lsem \varphi \rsem(\mathit{nw}, \sigma)
	&=\lsem \psi \wedge \forall x.\theta_1\rsem_{\mathcal{V}}(\mathit{nw}, \sigma) \\
	&= \lsem \psi \rsem_{\mathcal{V}}(\mathit{nw}, \sigma) \diamond \Val^\omega((\lsem \theta_1 \rsem_{\mathcal{V}\cup \{x\} }(\mathit{nw},\sigma [x \rightarrow i]))_{i \in \mathbb{N}}) \\
	&\stackrel{*}{=} \Val^\omega((\lsem \psi \rsem_{\mathcal{V}}(\mathit{nw}, \sigma) \diamond \lsem \theta_1 \rsem_{\mathcal{V}\cup \{x\} }(\mathit{nw},\sigma [x \rightarrow i]))_{i \in \mathbb{N}}) \\
	&= \Val^\omega((\lsem \psi \rsem_{\mathcal{V}\cup\{x\}}(\mathit{nw}, \sigma[x \rightarrow i]) \diamond \lsem \theta_1 \rsem_{\mathcal{V}\cup \{x\} }(\mathit{nw},\sigma [x \rightarrow i]))_{i \in \mathbb{N}}) \\
	&= \Val^\omega((\lsem \psi \wedge \theta_1 \rsem_{\mathcal{V}\cup \{x\} }(\mathit{nw},\sigma [x \rightarrow i]))_{i \in \mathbb{N}}) \\
	&= \lsem \forall x.(\psi \wedge \theta_1)\rsem_{\mathcal{V}}(\mathit{nw}, \sigma) \enspace .
\end{align*}
So $\varphi'=\forall x. \varphi_1$ is strongly-$\wedge$-restricted and $\lsem \varphi \rsem=\lsem \varphi' \rsem$.
If $\varphi$ is $\forall$-restricted, $\theta_1$ is almost boolean. In this case we can put directly $\varphi'=\forall x.(\psi \wedge \theta_1)$. Then $\varphi'$ is strongly-$\wedge$-restricted and $\forall$-restricted because $\psi$ and $\theta_1$ are almost boolean formulas.
\par
Now let $D$ be left-multiplicative.
Using the formulas $min(x) = \forall y. ( x \leq y)$ and $min(x) \rightarrow \psi = \neg min(x) \vee (min(x) \wedge \psi)$ it can be shown that
\begin{align*} \lsem \varphi \rsem%
	&= \lsem \psi \wedge \forall x.\theta_1\rsem \\
	&= \lsem \forall x.((min(x) \rightarrow \psi) \wedge \theta_1)\rsem\\
	&= \lsem \forall x.((\neg min(x) \wedge \theta_1) \vee (min(x) \wedge \psi \wedge \theta_1)) \rsem \enspace.
\end{align*}
Then $\varphi'=\forall x.((\neg min(x) \wedge \theta_1) \vee (min(x) \wedge \varphi_1))$ is strongly-$\wedge$-restricted since $min(x)$ is boolean. Furthermore,
$\lsem \varphi \rsem=\lsem \varphi' \rsem$.
If $\varphi$ is $\forall$-restricted, we can put directly $\varphi'=\forall x.((min(x) \rightarrow \psi) \wedge \theta_1)$. Then $\varphi'$ is strongly-$\wedge$-restricted and $\forall$-restricted because $min(x) \rightarrow \psi$ and $\theta_1$ are almost boolean formulas.
\end{proof}
If $D$ is a cc-$\omega$-valuation semiring, clearly
 almost boolean formulas can be written as disjunctions
 of conjunctions of boolean formulas or constants from $D$.
 Our proof of Theorem \ref{thm:restrict} $(b)$ shows the following corollary.
\begin{Folgerung}
 Let $D$ be a commutative cc-$\omega$-valuation semiring.
 Then for any formula $\varphi \in \mathit{MSO}(D, \mathit{NW}(\Sigma))$ there exists a formula $\varphi' \in \mathit{MSO}(D, \mathit{NW}(\Sigma))$ in which conjunctions occur
 only between boolean formulas and constants such that $\lsem \varphi \rsem =\lsem \varphi'\rsem$.
\end{Folgerung}
 This follows also from a slightly modified proof of Theorem \ref{main},
 but the present proof gives direct and efficient
 conversions of the formulas.
	\section{Characterization of Regular Series} \label{chapchar}
In this section, we give our second main result on the expressive equivalence of weighted stair Muller nested word automata and our different fragments of weighted MSO logic. 
\begin{Theorem}
\label{main}
	Let $D$ be a %
		regular $\omega$-pv-monoid and $S: \mathit{NW^\omega}(\Sigma)\rightarrow D$ a series.
	\begin{enumerate}%
	\item 	The following are equivalent:
		\begin{enumerate}%
			\item $S \text{ is regular}$.
			\item $S=\lsem \varphi \rsem \text{ for some syntactically restricted sentence }\varphi \text{ of } \\ \mathit{MSO}(D, \mathit{NW}(\Sigma))$.
		\end{enumerate}
	\item 	Let $D$ be left-distributive. Then the following are equivalent:
		\begin{enumerate}%
			\item $S \text{ is regular}$.
			\item $S=\lsem \varphi \rsem \text{ for some $\forall$-restricted and } \text{$\wedge$-restricted sentence } \varphi \text{ of } \\ \mathit{MSO}(D, \mathit{NW}(\Sigma))$.
		\end{enumerate}
	\item 	Let $D$ be cc-$\omega$-valuation semiring. Then the following are equivalent:
		\begin{enumerate}%
			\item $S \text{ is regular}$.
			\item $S=\lsem \varphi \rsem \text{ for some $\forall$-restricted and } \text{commutatively-$\wedge$-restricted sentence } \\ \varphi \text{ of } \mathit{MSO}(D, \mathit{NW}(\Sigma))$.
		\end{enumerate}
	\end{enumerate}
\end{Theorem}
\begin{proof}
\par 
\textbf{'$(i) \Rightarrow (ii)$':} We construct a syntactically %
restricted MSO-sentence simulating the given wsMNWA, thus showing all three statements. 
\par
\textbf{'$(ii) \Rightarrow (i)$':} %
By Theorem \ref{thm:restrict} we may assume $\varphi$ to be syntactically restricted. %
We prove the regularity of $\lsem \varphi \rsem$ by induction on the structure of $\varphi$ %
as follows.
If $\varphi$ is almost boolean, %
by Propositions \ref{almboolrsf}(a) and \ref{rsfreg}, $\lsem \varphi \rsem$ is regular. %
Next we have to prove that the regularity is preserved under the non-boolean operations.
We only sketch the ideas. Closure under disjunction %
follows from Proposition \ref{hom} and a union construction of automata. %
If $\varphi$ is a conjunction, the regularity of $\lsem \varphi \rsem$ 
follows from a product construction of automata. The regularity of $\lsem \exists x. \varphi\rsem$ and $\lsem \exists X. \varphi \rsem$ follows from Proposition \ref{hom}. For $\forall x. \varphi$, $\varphi$ is almost boolean. Then $\lsem \forall x.\varphi \rsem$ can also be shown to be regular. %
\end{proof} 

	\section{Conclusion} 
We have introduced a weighted automaton model for infinite nested
words and weighted MSO logics. We could show that under suitable
assumptions on the valuation monoids, %
two resp. three fragments of the weighted logics have the same
expressive power with efficient conversions into the smallest
fragment. Moreover, the weighted automata and our logic fragments
have the same expressive power. The valuation monoids
form very general weight structures; they model
long-time average and discounted computations of weights as well
as the classical complete semirings \cite{DG}. As in \cite{AM}, we considered
nested words possibly containing pending edges. We remark that
our results also hold similarly for finite nested words, and our conversions of the weighted logic formulas also work, similarly, for other discrete structures like trees, cf. \cite{DGMM}. 

It would be interesting to investigate decision problems for
weighted nested word automata, e.g., like done in \cite{CDH}, \cite{CDH2} 
for automata on words and with average or discounted computations
of weights.%
\bibliography{sdbib}{}
\bibliographystyle{splncs03}
\end{document}